\documentclass[letterpaper,english,aps,prb,floatfix,showpacs,twocolumn,amsfonts,amssymb,superscriptaddress,longbibliography]{revtex4-2}

\usepackage{balance}
\usepackage[T1]{fontenc}
\usepackage[utf8]{inputenc}
\usepackage[english]{babel}
\usepackage{graphics}
\usepackage{amssymb}
\usepackage{amsmath, esint}
\usepackage{overpic}
\usepackage[toc]{appendix}
\usepackage{bbold}

\usepackage{natbib}

\usepackage{xcolor}
\usepackage{soul}
\usepackage{commath}
\usepackage{tikz}
\usepackage{physics}

\usepackage{hyperref}

\newcommand{\be}{\begin{equation}}
\newcommand{\ee}{\end{equation}}
\newcommand{\bspl}{\begin{split}}
\newcommand{\espl}{\end{split}}
\newcommand{\bea}{\begin{eqnarray}}
\newcommand{\eea}{\end{eqnarray}}

\newcommand{\angstrom}{\mbox{\normalfont\AA}}

\def\bq{{\bf q}}

\def\nn{\nonumber}
\def\lb{\label}
\def\pref#1{(\ref{#1})}

\newcount\bozza \bozza=0
\ifnum\bozza=1
\newdimen\shift \shift=-2truecm
\def\lb#1{%
{\label{#1}\rlap{\kern\shift{$\scriptstyle#1$}}}}
\else\def\lb#1{\label{#1}} \fi

\definecolor{darkred}{rgb}{0.55, 0.0, 0.0}
\definecolor{darkpowderblue}{rgb}{0.0, 0.2, 0.6}

\usetikzlibrary{shapes.arrows, fadings}
\tikzfading[name=fadeleft,
  left color=transparent!80, right color=transparent!0]

\begin{document}
\title{Ghost Josephson plasmon in bilayer superconductors}
\author{Niccolò Sellati}
\email{niccolo.sellati@uniroma1.it}
\affiliation{Department of Physics and ISC-CNR, ``Sapienza'' University of Rome, P.le A. Moro 5, 00185 Rome, Italy}
\author{Lara Benfatto}
\email{lara.benfatto@roma1.infn.it}
\affiliation{Department of Physics and ISC-CNR, ``Sapienza'' University of Rome, P.le A. Moro 5, 00185 Rome, Italy}

\begin{abstract}
The experimental measurement of collective charge fluctuations in metals and superconductors is a preferential tool to benchmark fundamental interactions in solids. Recent experiments in multicomponent systems, from superconducting layered cuprates to multiband metals, highlighted striking effects due to the interplay between different degrees of freedom. In this paper we provide a physical explanation for the existence of a "ghost" Josephson plasmon in bilayer superconductors, layered systems with two layers per unit cells that interact with two different Josephson couplings. We show that one of the two plasmons that emerge after the breaking of the translational symmetry along the out-of-plane direction is connected to counterflowing current fluctuations polarized perpendicularly to the planes. This effect makes it a staggered mode that is virtually transverse at small out-of-plane momenta $q_c$, explaining why it is hidden in the density response at small $q_c$. Our work offers an additional perspective on the understanding of collective excitations in systems with multiple intertwined degrees of freedom.
\end{abstract}
\date{\today}

\maketitle

\section{Introduction}
The propagation of electromagnetic (e.m.) waves in metals gives rise to two hybrid light-matter modes, named plasmons or plasma-polaritons depending on their polarization \cite{maier}. Plasmons can be easily understood semiclassically in a jellium model as the longitudinal oscillations of the electric field due to the relative accumulation of electronic charge density as compared to a uniform ionic background, in a capacitor-like picture. Instead, polaritons identify transverse e.m.\ waves propagating in a polar medium, where the light-induced polarization of the matter gives rise to a frequency-dependent dielectric function. In the limit of vanishing momentum $\textbf q \to 0$ these excitations are degenerate, approaching the so-called plasma frequency $\omega_p^2=4\pi e^2 n/m^*$, with $n$ the electron density and $m^*$ the effective electronic mass of the metal \cite{maier}. The small-momentum regime can be probed in optical experiments, where the external drive is an e.m.\ wave with momentum that is typically negligible as compared to the lattice momenta. At finite $\textbf q$ the dispersion of the transverse polariton rapidly approaches the asymptotic light dispersion as $\omega^2(\textbf q)\simeq \omega_p^2+c^2|\textbf q|^2/\varepsilon_\infty$, with $c$ the light velocity and $\varepsilon_\infty$ the background dielectric constant, while the longitudinal plasmon disperses with velocity $v_p$ of the order of the Fermi velocity, as $\omega^2(\textbf q)\simeq \omega^2_p+v^2_p|\textbf q|^2$. The latter can be measured with experimental techniques like resonant inelastic X-ray scattering (RIXS) \cite{vanderbrink_review} or electron energy loss spectroscopy (EELS) \cite{abajo_review}, that are able to probe density fluctuations at finite momenta.
In other words, while optics essentially probes the transverse current-current correlation functions of the metal \cite{basov-review-metals}, RIXS and EELS probe the density-density correlation function \cite{abajo_review,vanderbrink_review}. This paradigm has been well-confirmed in both ordinary and correlated single-band metals, with the remarkable difference that probing collective modes in the latter also provides a direct access to electronic interactions \cite{basov-review-polaritons}. 
A typical example is the observation of strongly damped plasma modes in the metallic and superconducting (SC) phase of cuprates via EELS and RIXS spectroscopies, which prompted an intense investigation on the nature of charge fluctuations in these complex materials \cite{fink_prb89,fink_prb91,markiewicz_prb08,greco_prb16,abbamonte_scipost17,lee_rixs_nature18,mitrano_pnas18,mitrano_prx19,liu_rixs_npjqm20,greven_prx20,zhou_prl20,huang_rixs_prb22,hepting_prl22,phillips_prb22,abbamonte_natcomm23,abbamonte_nature23, hepting_rixs_prb24,zhou_cm24,fink_cm24,abbamonte_fink_eels_review,abbamonte_prb24}. 

On the other hand, even for weakly correlated systems the previous picture can considerably change when {\em multiple} degrees of freedom are at play simultaneously, leading to interesting effects that can still be understood within a Fermi-liquid-like paradigm. An interesting example is the case of multiband metals, where several pockets coexist at the Fermi level. In this case, in addition to ordinary plasmons, acoustic-like modes can also emerge in the presence of a large anisotropy between the plasma energy scales of the various bands, as a result, e.g., of very different effective masses. 
The physical mechanism behind this effect,  originally discussed by Pines in 1956 \cite{pines_demon1956}, 
is that light particles are able to completely screen out the Coulomb field of the heavy particles at finite momentum, leading to a sound-like mode, the "Pines' demon", instead of a gapped plasmon. This effect can be captured within an RPA computation of the density response, as recently discussed in connection to the observation of an acoustic-like branch in $\text{Sr}_2\text{RuO}_4$ \cite{abbamonte_nature23}.\\
A somehow analogous effect manifests in layered metals, where the anisotropy of charge fluctuations within and in-between the weakly-coupled planes leads to acoustic-like plasmon branches for finite values of the out-of-plane momentum. Layered models provide a paradigmatic description of SC cuprates, whose plasma modes have recently attracted considerable interest in connection not only to direct measurements of the density response, but also in relation to experiments with strong THz light pulses \cite{cavalleri_natphys16,cavalleri_science18,averitt_pnas19,cavalleri_prb22,averitt_prb23,shimano_prb23,cavalleri_prx22,shimano_prb23,cavalleri_nmat14,wangNL_science24}. Indeed, since the metallic plasmons of the normal phase are mapped into "Josephson" plasmons in the SC phase, which are intrinsically anharmonic modes, intense THz electric fields can be used to drive them in a nonlinear regime \cite{nori_natphys06,nori_review10,demler_prb20,gabriele_natcomm21,demler_commphys22,fiore_prb24}.\\
Some layered materials can have more complex structures, leading to the coexistence of even more plasma scales that affect the plasma response. A prototypical example are the bilayer metals, crystals with two layers per unit cell. In these systems, the difference in the out-of-plane hopping between the intrabilayer and interbilayer spacings translates in a marked anisotropy between out-of-plane plasma frequency scales.
In general, the doubling of the layers per unit cell results in a doubling of the plasmon branches, as observed, e.g., in linear optical spectroscopy on the cuprate $\text{YBa}_2\text{Cu}_3\text{O}_7$ (YBCO) below the critical temperature $T_c$, where the quasiparticle continuum is gapped and the modes become undamped \cite{cavalleri_nmat14,vandermarel96,homes_prl93,bernhard_prl11}. 
On the other hand, recent calculations of the density response at RPA level \cite{hepting_rixs_prb24,stoof_prb24,yamase_cm24} show that the lower plasmon branch is \textit{hidden} in the density response for small out-of-plane momenta $q_c$. This result has been discussed in connection with recent RIXS measurements on the bilayer $\text Y_{0.85}\text{Ca}_{0.15}\text{Ba}_2\text{Cu}_3\text{O}_7$ (Ca-YBCO) \cite{hepting_rixs_prb24}, where only one plasmon branch is observed. Besides the issue of the correct comparison with experiments, that is still debated \cite{yamase_cm24},  
no theoretical insight has been provided up to now to the physical reason \textit{why} one "ghost" plasmon exists in a bilayer superconductor. In particular, it would be interesting to explore possible analogies with the acoustic-like Pines' demon of multiband metals we mentioned above, that has been argued to be linked to out-of-phase density fluctuations in the different Fermi pockets \cite{abbamonte_nature23}. \\
In this work we show that the lower Josephson plasmon in bilayer superconductors is similarly associated with opposite-phase density oscillations between the two layers of each unit cell, but its behavior as a ghost at low $q_c$ has a completely different origin. We take advantage of a recent theoretical scheme implemented to study plasma waves in layered superconductors via the fluctuations of the phase of the SC order parameter \cite{benfatto_prb01,benfatto_prb04,gabriele_prr22,sellati_prb23,fiore_prb24,sellati_nano24}, that allows us to obtain an analytical expression for the density-density response function that can be mapped in the results obtained previously in the normal state \cite{hepting_rixs_prb24,yamase_cm24}. In addition, we are able to connect its behavior to the polarization of the microscopic currents associated with the plasma modes, and in particular we show that the lower Josephson plasmon is invisible in the density response because it is a \textit{transverse} excitation at small $q_c$. 
This apparently unexpected behavior of the polarization of the mode can be understood with symmetry-breaking considerations on the single-layer crystal: when the translation symmetry along the out-of-plane $c$-axis is broken, the dispersion of the single-layer Josephson plasmon at the edge of the Brillouin zone backfolds at $q_c=0$. The new mode originating from the symmetry breaking, the lower Josephson plasmon, retains the polarization of the original mode with a "reminiscent" large out-of-plane momentum. The plasmon thus becomes longitudinal only for large enough $q_c$, where it becomes visible in the density response. Because of this mechanism, 
the spectral weight of the density-density response function is not $2\pi/d$ periodic. 
Moreover, one observes that the dispersion of the lower Josephson plasmon is approximately acoustic as a function of the in-plane momentum. Interestingly, the association of opposite-phase oscillations with an acoustic mode also holds for the Pines' demon observed in Ref.\ \cite{abbamonte_nature23}, calling for a possible generalization of the connection between the two features. 
For what concerns the RIXS measurements of 
Ref.\ \cite{hepting_rixs_prb24}, if one uses the experimental value of the out-of-plane momentum $q_c=1.8\pi/d$, the most natural interpretation is that the measured mode is the lower Josephson plasmon, as recently argued in Ref.\ \cite{yamase_cm24}, with the upper mode probably being overdamped at higher frequencies.
On more general grounds, our work offers yet another perspective on the complex nature of collective modes in systems with multiple interacting degrees of freedom,
and highlights the physical mechanisms responsible for the dispersion of the collective modes and their coupling to physical probes. 
\section{Plasma mode response in single-layer superconductors}
\subsection{Plasma modes in single-layer systems}
Before studying the case of bilayer superconductors it is instructive to review briefly the procedure to compute the generalized plasma modes and the density-density response function in single-layer superconductors, as e.g., the cuprate LSCO. 
As mentioned in the Introduction, plasma modes can be efficiently described in the SC state by means of the degrees of freedom linked to the fluctuations of the phase of the complex order parameter. More specifically, if one derives the quantum action for the SC phase by means of a path-integral approach, the propagator of the phase mode directly identifies the dispersion of the plasma mode \cite{nagaosa,dassarma_prl90,dassarma_prb91,dassarma_prb95,depalo_prb99,randeria_prb00,benfatto_prb01,benfatto_prb04,millis_prr20,gabriele_prr22,sellati_prb23,fiore_prb24,sellati_nano24}, without the need to expand the fermionic bubbles appearing in the density response computed at RPA level in the metallic state \cite{abbamonte_nature23,anderson_pr58}. In other words, the phase action, that only depends on the time and space derivatives of the SC phase, automatically implements the frequency and momentum structure of the plasma modes, with considerable simplifications in the formalism for layered systems. In addition, as we will see below, the phase action allows one to derive in a compact and elegant way the polarizations of the plasma modes. Even though our results are formally analogous to the RPA derivation of the density response for the metallic state obtained in previous works \cite{hepting_rixs_prb24,yamase_cm24,stoof_prb24}, they allow us for a transparent physical interpretation of the emerging ghost plasma mode in the bilayer case.\\
To understand how plasma modes emerge from a phase-only action, we start from the Gaussian action of a neutral, layered superconductors, that reads \cite{depalo_prb99,randeria_prb00,benfatto_prb01,millis_prr20,gabriele_prr22,benfatto_prb04,sellati_prb23,sellati_nano24}
\begin{align}\lb{sthetaSL}
    S_G[\theta]=\frac{d}{8}\sum_{q}[\kappa_0\Omega_m^2+D_{ab}k_{ab}^2+D_ck_c^2]|\theta(q)|^2,
\end{align}
where $q=(i\Omega_m,\textbf q)$ is the imaginary-time 4-momentum, with $\Omega_m=2\pi mT$ the bosonic Matsubara frequencies and $\textbf q$ the 3-momentum. The in-plane components of the momentum are in $k_{a,b}=2/a\sin(q_{a,b}a/2)$, with $k_{ab}^2=k_a^2+k_b^2$ and $a$ the in-plane lattice constant, and the out-of-plane component is in $k_c=2/d\sin(q_cd/2)$, with $d$ the distance between SC planes. The coefficient $\kappa_0$ is the bare compressibility while $D_{ab}$ and $D_c$ are respectively the in-plane and out-of-plane superfluid stiffnesses, which reduce to the ratio between the superfluid density and the mass of the charge carriers in the continuous isotropic limit. As anticipated, the propagator of the Gaussian phase action identifies the phase mode once the analytical continuation $i\Omega_m\to \omega +i0^+$ is performed: one then finds from Eq.\ \pref{sthetaSL} a sound-like mode, as expected since the SC phase is the Goldstone mode of the broken $U(1)$ symmetry in the SC state \cite{nagaosa}. 
To simplify the notation, in the following we will always assume the in-plane momentum to be along the $a$-direction ($q_b=0$) without lack of generality.\\
The promotion of the sound mode to the plasmon occurs thanks to the coupling of the SC phase to the e.m.\ interactions \cite{anderson_pr58}. This can be achieved by performing in Eq.\ \eqref{sthetaSL} the minimal-coupling substitution \cite{nagaosa},
\begin{align}\lb{mincoupSL}
    \Omega_m\theta(q)\to\Omega_m\theta(q)-2e\phi(q),\nn\\
    i\textbf k \theta(q)\to i \textbf k \theta(q)+\frac{2e}{c}\textbf A(q),
\end{align}
and by adding the action for the free e.m.\ field,
\begin{align}\lb{freeemSL}
    S_\text{e.m.}[\phi,\textbf A]&=\frac{\varepsilon_\infty d}{8\pi}\sum_q\bigg[\frac{\Omega_m^2}{c^2}|\textbf A(q)|^2+\frac{1}{\varepsilon_\infty}|\textbf k\times \textbf A(q)|^2\nn\\
    &-|\textbf k|^2|\phi(q)|^2+\frac{2i\Omega_m}{c}\phi(q)\textbf k \cdot\textbf A(-q)\bigg],
\end{align}
where $\phi$ and $\textbf A$ are the e.m.\ scalar and vector potential respectively, $-e$ is the charge of the electron, $c$ is the light velocity and $\varepsilon_\infty$ is the background dielectric constant. The effect of the minimal substitution \pref{mincoupSL} is twofold: (i) one obtains the screening of the e.m.\ interaction by modification of the e.m.\ action in vacuum Eq.\ \eqref{freeemSL}, and (ii) one couples the SC phase degrees of freedom to the e.m.\ field via the term: 
%
\begin{align}\lb{coupactSL}
    S_\text c&[\theta,\phi,\textbf A]=\frac{ed}{4}\sum_q\big[2\kappa_0\Omega_m\theta(q)\phi(-q)\nn\\
    &+\frac{2i}{c}\theta(q)\big(D_{ab}k_{a}\text A_{ab}(-q)+D_{c}k_{c}\text A_{c}(-q)\big)\big].
\end{align}
In the isotropic case, where $D_{ab}=D_c\equiv D_s$, the second line of Eq.\ \pref{coupactSL} reduces to $D_s\theta(q) (\textbf k\cdot \textbf A(-q))$. The coupling between SC phase and vector potential thus vanishes in the Coulomb gauge ($\textbf k\cdot \textbf A$=0) and one can retain only the coupling of $\theta$ to the scalar potential. In the anisotropic case this is in general not possible. However, as discussed in Refs.\ \cite{gabriele_prr22,sellati_prb23}, neglecting the coupling between $\theta$ and $\textbf A$ is equivalent to neglect retardation effects, and it turns out to be a quantitatively good approximation for the momenta probed by RIXS \cite{vanderbrink_review} and EELS \cite{abajo_review}. Because of this we will retain only the first term of Eq.\ \pref{coupactSL}. The total action for the phase and scalar potential is then given by:
\begin{align}\lb{RPAactSL}
    S_\text{RPA}&[\theta,\phi]= S_G[\theta]-\frac{\varepsilon_\infty d}{8\pi}\sum_q\big[(|\textbf k|^2+ k_\text{TF}^2)|\phi(q)|^2\big]\nn\\
    &+\frac{ed}{4}\sum_q\big[\kappa_0\Omega_m\big(\theta(q)\phi(-q)-\theta(-q)\phi(q)\big)\big],
\end{align}
where $k_\text{TF}^2=4\pi e^2 \kappa_0/\varepsilon_\infty$ is the Thomas-Fermi wavevector, that as usual enters as a cut-off for the long-wavelength part of the screened Coulomb potential. By integrating out the e.m.\ scalar potential one dresses the sound-like SC phase mode with the Coulomb interaction, obtaining the plasma mode:
\begin{align}\lb{RPAdressSL}
    S_\text{RPA}[\theta]=\frac{\varepsilon_\infty d}{32\pi e^2}\sum_q \frac{|\textbf k|^2}{1+\alpha|\textbf k|^2}\bigg[{\Omega_m^2}+\omega_\text{RPA}^2(\textbf q)\bigg]|\theta(q)|^2,
\end{align}
where $\alpha=\lambda_D^2=1/k_\text{TF}^2$ is the square of the Debye screening length and where we defined 
\begin{align}\lb{RPAdispSL2}
    \omega^2_\text{RPA}(\textbf q)=\omega_L^2(\textbf q)(1+\alpha|\textbf k|^2),
\end{align}
with
\begin{align}\lb{RPAdispSL}
    \omega_L^2(\textbf q)=\omega_{ab}^2\frac{k_a^2}{|\textbf k|^2}+\omega_c^2\frac{k_c^2}{|\textbf k|^2},
\end{align}
$\omega_{ab}^2=4\pi e^2 D_{ab}/\varepsilon_\infty$ and $\omega_{c}^2=4\pi e^2 D_{c}/\varepsilon_\infty$ being the in-plane and out-of-plane plasma frequencies respectively. As anticipated, the pole of the phase propagator given by the Gaussian action Eq.\ \eqref{RPAdressSL} describes a longitudinal Josephson plasmon with dispersion relation given by $\omega_\text{RPA}(\textbf q)$ in Eq.\ \pref{RPAdispSL2}. The same plasmon dispersion could be obtained by RPA resummation of the density fluctuations \cite{randeria_prb00,benfatto_prb01,millis_prr20,benfatto_prb04,dassarma_prl90,dassarma_prb91,dassarma_prb95}, that consists in replacing the bare compressibility $\kappa_0$ in Eq.\ \eqref{sthetaSL} with $\kappa_0/(1+V(\textbf q))$, $V(\textbf q)=4\pi e^2/\epsilon_\infty |\textbf k|^2$ being the Coulomb potential. In typical layered superconductors the effects on the dispersion of a finite compressibility, i.e.\ a finite $\alpha$, are negligible at small momenta when compared to the dispersive behavior given by anisotropy, i.e.\ $\omega_\text{RPA}(\textbf q)\simeq \omega_L(\textbf q)$. Such corrections are, however, important for higher momenta probed by RIXS and EELS. In Fig.\ \ref{fig1}(a) we show the dispersion Eq.\ \eqref{RPAdispSL2} of the Josephson plasmon as a function of $q_a$ for fixed $q_c$.
\begin{figure}[t!]
    \centering
    \includegraphics[width=0.49\textwidth,keepaspectratio]{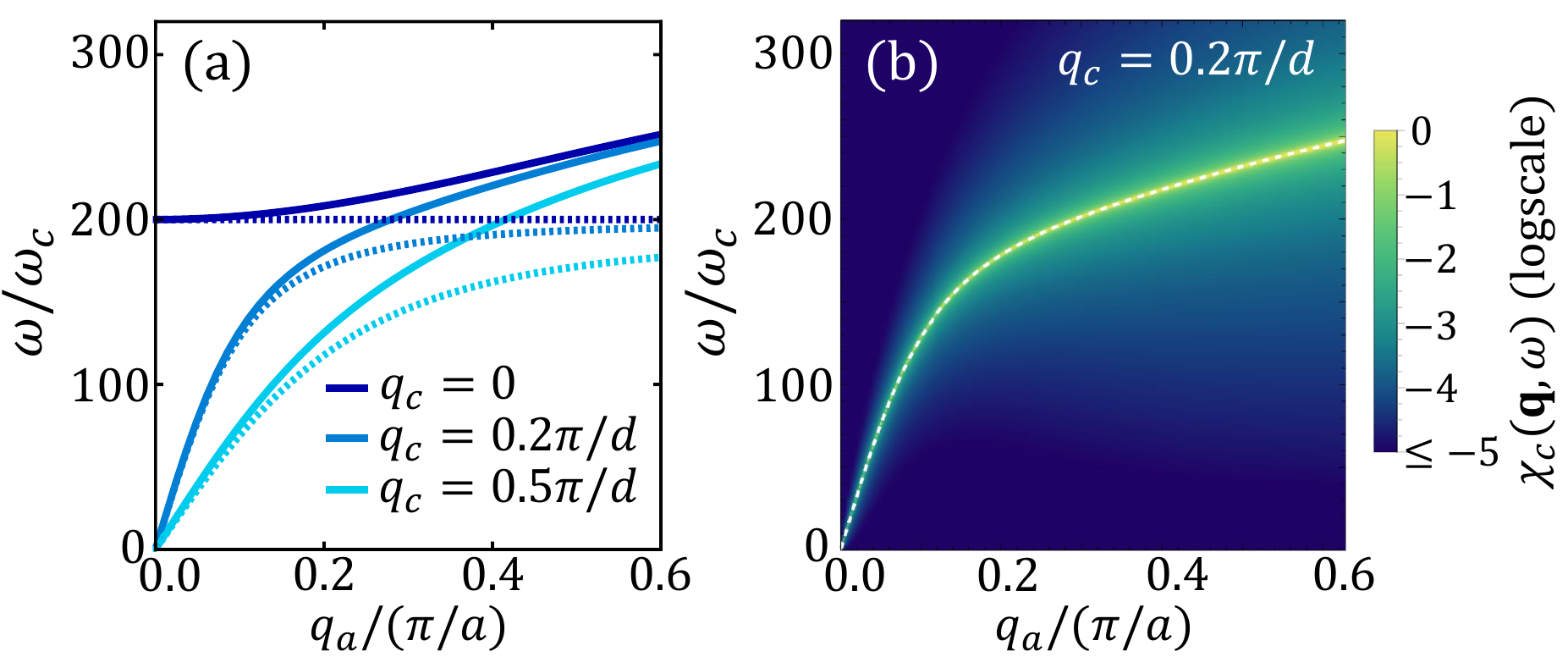}
    \caption{(a) Josephson plasma mode dispersion $\omega_\text{RPA}(\textbf q)$ according to Eq.\ \eqref{RPAdispSL2} as a function of $q_a$ for some fixed $q_c$. Dashed lines are $\omega_L(\textbf q)$ according to Eq.\ \eqref{RPAdispSL}. (b) Intensity map of $\chi_c(\textbf q,\omega)$ of a single-layer superconductor according to Eq.\ \eqref{measquantSL} as a function of $q_a$ for fixed $q_c=0.2\pi/d$. The response function is normalized to its maximum value. White dashed line represents $\omega_\text{RPA}(\textbf q)$. In the plots, $\omega_{ab}/\omega_c=200$, $a/d=0.6$, $\alpha/d^2=0.08$ and $\gamma/\omega_c=0.05$. }
    \lb{fig1}
\end{figure}
\subsection{RPA density-density response function in single-layer systems}
Once clarified how the spectrum of the plasmon is described by the phase degrees of freedom, we can compute the density-density response of the superconductor within the same formalism of the quantum action. As discussed in Refs.\ \cite{cea_prb16,gabriele_prr22}, we can introduce into the microscopic Hamiltonian an auxiliary scalar field $\delta\phi(q)$ coupled to the density operator and then compute the functional derivative of the effective action with respect to it to obtain the density-density response function, 
\begin{align}\lb{densdensdef}
    \chi_{\rho\rho}(q)=\frac{1}{d}\frac{\partial^2 S[\delta\phi]}{\partial \delta\phi(q) \partial \delta\phi(-q)}.
\end{align}
The auxiliary field plays the role of an external scalar potential, so its coupling to the phase degrees of freedom can be obtained by an extension of the minimal-coupling substitution of Eq.\ \eqref{mincoupSL}:
\begin{align}\lb{mincoupSL2}
    \Omega_m\theta(q)\to\Omega_m\theta(q)-2e\phi(q)-2e\delta\phi(q).
\end{align}
We then integrate out the e.m.\ scalar potential and the SC phase to find the Gaussian effective action $S[\delta\phi]$. With straightforward calculations, detailed in Appendix A, one finds:
\begin{align}\lb{densdensfuncSL}
    \chi_{\rho\rho}(q)=-\frac{\varepsilon_\infty}{4\pi}\frac{|\textbf k|^2\omega_L^2(\textbf q)}{\omega_\text{RPA}^2(\textbf q)-(\omega+i0^+)^2},
\end{align}
where we performed the analytic continuation $i\Omega_m\to\omega+i0^+$. This expression is formally equivalent to the RPA result in the metallic state. In experiments, the measured quantity is proportional to the total charge response $\chi_{c}(\textbf q,\omega)$, that is the imaginary part of the response function at positive frequencies \cite{vanderbrink_review,abajo_review,abbamonte_fink_eels_review}, which from Eq.\ \eqref{densdensfuncSL} reads
\begin{align}\lb{measquantSL}
    \chi_c(\textbf q,\omega)=-\text{Im}(\chi_{\rho\rho})=W(\textbf q)\delta\big(\omega-\omega_\text{RPA}(\textbf q)\big),
\end{align}
where the spectral weight of the delta-function is
\begin{align}\lb{weightSL}
    W(\textbf q)=\frac{\varepsilon_\infty}{8}\frac{|\textbf k|^2\omega_L(\textbf q)}{\sqrt{1+\alpha|\textbf k|^2}}.
\end{align}
%
Eq.\ \eqref{measquantSL} is peaked at $\omega_\text{RPA}(\textbf q)=\omega_L(\textbf q)\sqrt{1+\alpha|\textbf k|^2}$ \cite{zhou_prl20,zhou_cm24}, that is the dispersion of the longitudinal Josephson plasmon as in Eq.\ \eqref{RPAdispSL2}. In Fig.\ \ref{fig1}(b) we show $\chi_c(\textbf q,\omega)$ as a function of $q_a$ for fixed $q_c$, where we introduced a phenomenological damping parameter $\gamma$ while performing the analytic continuation $i\Omega_m\to\omega+i\gamma$. 
The spectral weight of the plasmon scales as $|\textbf k|^2$, thus vanishing as $|\textbf q|^2$ for $\textbf q\to 0$, as expected by the gauge-invariance requirement that the dynamical density response vanishes at long-wavelength \cite{cea_prb16,devereaux_prb95}. It is worth noting that in the usual RPA expression for the density response in the metallic state such $|\textbf q|^2$ scaling follows from the expansion of the bare density susceptibility $\chi^0(\textbf q,\omega)$ \cite{coleman}:
\begin{align}\lb{densdensmet}
    \chi^\text{metal}_{\rho\rho}(q)=\frac{\chi^0(\bq,\omega)}{1+V(\bq)\chi^0(\bq,\omega)},
    \end{align}
Considering, e.g., the isotropic case, by expanding $\chi^0(\textbf q,\omega)\approx -n |\textbf q|^2/(m\omega^2)$ in the limit $\omega /|\textbf q|\gg 1$ one recovers the isotropic analogous of Eq.\ \eqref{densdensfuncSL}. Instead, while computing the density response below $T_c$ via the phase mode, the correct momentum dependence of $\chi_{\rho\rho}(q)$ is automatically obtained once the phase mode is integrated out, as we have shown. This is a consequence of the fact that the phase action Eq.\ \pref{sthetaSL} is obtained from an expansion in time and space gradients of the phase, making the leading powers in frequency and momentum explicit. This finding will turn out to be crucial in the bilayer case to obtain analytically transparent expressions for the spectral weights of the two different plasma modes. 

\section{Plasma mode response in bilayer superconductors}
\subsection{Plasma modes and density response in bilayer systems}
We now generalize the framework developed in the previous sections to the case of bilayer superconductors as, e.g., the cuprate YBCO, which are systems with two SC layers per unit cell. In the following we keep the label $d$ for the lattice constant along the $c$-axis, whereas we label with $d_1$ the spacing between the SC planes within a same bilayer unit (intrabilayer) and with $d_2$ the distance between subsequent planes in two different unit cells (interbilayer), see Fig.\ \ref{fig2}(a) for the notation followed in this section. Due to the broken translational symmetry along the $c$-axis and the different nature of the insulating layers in-between the SC sheets, 
the Josephson-like coupling that causes the tunneling of Cooper pairs in the intrabilayer spacing is different from the coupling in the interbilayer spacing \cite{vandermarel96,vandermarel_prb01,fazio_review01,sellati_prb23}. To take this into account we introduce two different scales for the out-of-plane superfluid stiffness, $D_{c1}$ and $D_{c2}$ respectively. This immediately implies the existence of two Josephson plasma resonances, at frequencies $\omega_{c1}^2=4\pi e^2D_{c1}/\varepsilon_\infty$ and $\omega_{c2}^2=4\pi e^2D_{c2}/\varepsilon_\infty$. In the following we will consider $\omega_{c1}>\omega_{c2}$.\\
To keep the discussion of the physical results as transparent as possible, we report in Appendix B all the details of the derivation and we summarize here the main differences with respect to the single-layer case, discussed in the previous Section. First of all, since the phase fluctuations in the two SC planes are generically different, we define a vector containing the two modes $\boldsymbol{\theta}(q)=\big(\theta_1(q),\theta_2(q)\big)^T$, where $\theta_\lambda$ refers to the fluctuations in the $\lambda$th layer. The phase propagator found from the generalization of the single-layer action Eq.\ \pref{sthetaSL} is then promoted to a $2\times 2$ matrix. Analogously, the 
scalar potential is represented by a vector $\boldsymbol{\phi}(q)=\big(\phi_1(q),\phi_2(q)\big)^T$, and it can be introduced with the minimal-coupling substitution, generalizing Eq.\ \eqref{mincoupSL}, with a careful definition of the fields on the lattice and of the Fourier transform on the discrete $c$-direction \cite{homann_prr20,homann_prb21,sellati_prb23}. As in the single-layer case, the integration of the scalar potential promotes the phase modes to plasmons, and one obtains  the generalization of Eq.\ \eqref{RPAdressSL} to the bilayer case as
\begin{align}\lb{RPAdressBL}
    S_\text{RPA}[\boldsymbol\theta]&=\frac{\varepsilon_\infty}{32\pi e^2}\frac{d}{2}\sum_q\boldsymbol\theta^T(-q)\nn\\
    &\times\hat{\mathcal K}^2\big[\Omega_m^2(\hat{\mathbb1}+\alpha\hat{\mathcal K}^2)^{-1}+\hat\Omega_L^2(\textbf q)\big]\boldsymbol\theta(q),
\end{align}
where $\hat{\mathcal K}^2=k_a^2\hat{\mathbb1}+\hat{\mathcal K}_c^\dagger\hat{\mathcal K}_c$ and
\begin{align}\lb{RPAdispBL}
    \hat\Omega_L^2(\textbf q)=(\hat{\mathcal K}^2)^{-1}(\omega_{ab}^2k_a^2\hat{\mathbb1}+\hat{\mathcal K}_c^\dagger\hat\Omega_c^2\hat{\mathcal K}_c).
\end{align}
As one can see, Eqs.\ \eqref{RPAdressBL} and \eqref{RPAdispBL} have the same formal structure of Eqs.\ \eqref{RPAdressSL} and \eqref{RPAdispSL} respectively, with the $2\times 2$ matrices $\hat{\mathcal K}^2$, $\hat{\mathcal K}_c$ and $\hat{\Omega}_c^2$ playing the analogous role of the single-layer $|\textbf k|^2$, $k_c$ and $\omega_c^2$ \cite{sellati_prb23,fiore_prb24}, with
\begin{align}\lb{kmatrix}
    \hat{\mathcal K}_c=i\sqrt{\frac{2}{d}}\begin{pmatrix}
        \frac{e^{-iq_cd_1/2}}{\sqrt{d_1}} & -\frac{e^{iq_cd_1/2}}{\sqrt{d_1}}\\ 
        -\frac{e^{iq_cd_2/2}}{\sqrt{d_2}} & \frac{e^{-iq_cd_2/2}}{\sqrt{d_2}} 
    \end{pmatrix},
\end{align}
and
\begin{align}\lb{plasmamatrix}
    \hat{\Omega}_c^2=\begin{pmatrix}
        \omega_{c1}^2 & 0\\ 
        0 & \omega_{c2}^2 
    \end{pmatrix}.
\end{align}
The dispersions of the two Josephson plasmons of the bilayer superconductor can be found by solving the secular problem set by the action \pref{RPAdispBL}, i.e.\
\begin{align}\lb{modesBL}
\text{det}[\Omega_m^2\hat{\mathbb1}+\hat\Omega_L^2(\textbf q)(\hat{\mathbb1}+\alpha\hat{\mathcal K}^2)]=0,
\end{align}
where det is the determinant. Even though this equation can be solved analytically, we do not report here the rather lengthy expressions for the two modes. In the following we denote the two dispersions with $\omega_{+}(\textbf q)$ for the upper and $\omega_{-}(\textbf q)$ for the lower Josephson plasmons, shown in Fig.\ \ref{fig2}(b) as function of $q_a$ for some fixed $q_c$. 
%
\begin{figure}[t!]
    \centering
    \includegraphics[width=0.48\textwidth,keepaspectratio]{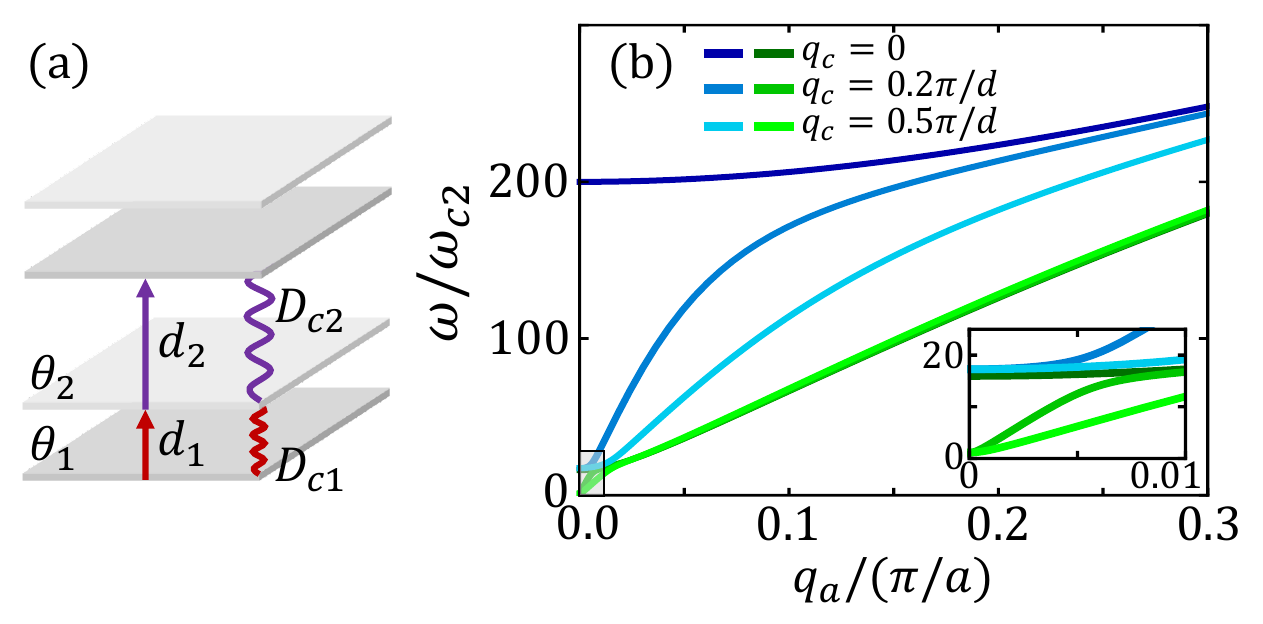}
    \caption{(a) Schematic representation of two subsequent bilayer units. The intrabilayer (red) and interbilayer (purple) Josephson-like interactions between planes that define two different superfluid stiffnesses are sketched as wavy lines. (b) Josephson plasma modes dispersions $\omega_{+}(\textbf q)$ (blue) and $\omega_{-}(\textbf q)$ (green) as function of $q_a$ for some fixed $q_c$. (inset) zoom on the gray shaded region of small momenta. In the plot we used $\omega_{ab}/\omega_{c2}=200$, $\omega_{c1}/\omega_{c2}=12$, $d_1/d=0.3$, $a/d=0.35$, $\alpha/d^2=0.08$, as appropriate for typical bilayer cuprates.}
    \lb{fig2}
\end{figure}
Once the plasma modes are determined, the computation of the density response closely follows the steps outlined for the single-layer, as detailed in Appendix B. More specifically, one introduces through the minimal-coupling two auxiliary fields $\delta\boldsymbol\phi(q)=\big(\delta\phi_1(q),\delta\phi_2(q)\big)^T$ coupled to the phase modes. After integration of the scalar potential and of the SC phase fluctuations one finds the Gaussian effective action $S_\text{RPA}[\delta\boldsymbol{\phi}]$, so that with a straightforward generalization of the definition Eq.\ \eqref{densdensdef}, one can obtain the density-density response function 
\begin{align}\lb{densdensfuncBL1}
    \hat\chi_{\rho\rho}(q)=-\frac{\varepsilon_\infty}{8\pi}\hat{\mathcal{K}}^2\big[\Omega_m^2\hat{\mathbb1}+\hat{\Omega}_L^2(\textbf q)(\hat{\mathbb1}+\alpha\hat{\mathcal{K}}^2)\big]^{-1}\hat{\Omega}_L^2(\textbf q).       
\end{align}
This is a $2\times 2$ matrix, accounting for the relative fluctuations of the density in each layer. Once again this expression is formally equivalent to the single-layer density-density response function in Eq.\ \eqref{densdensfuncSL}. One can immediately notice that the secular equation \eqref{modesBL} appears explicitly in the expression for the response function, and we can thus manipulate Eq.\ \eqref{densdensfuncBL1} using the identity $(\hat{\mathbb1}+\hat{B})^{-1}=\frac{\hat{\mathbb1}+\hat B^{-1}\det \hat B}{1+\Tr\hat B+\det\hat B}$, valid for a generic $2\times 2$ matrix $\hat B$ and where $\Tr$ is the trace, to make the dispersions of the two Josephson modes appear explicitly:
%
\begin{align}\lb{densdensfuncBL2}
    \hat\chi_{\rho\rho}&(q)=-\frac{\varepsilon_\infty}{8\pi}\frac{1}{\big(\Omega_m^2+\omega_+^2(\textbf q)\big)\big(\Omega_m^2+\omega_-^2(\textbf q)\big)}\nn\\
    &\times\bigg[k_a^2\big(\Omega_m^2\omega_{ab}^2\hat{\mathbb1}+\omega_+^2(\textbf q)\omega_-^2(\textbf q)(\hat{\mathbb1}+\alpha\hat{\mathcal{K}}^2)^{-1}\big)\nn\\
    &+\hat{\mathcal K}_c^\dagger\big(\Omega_m^2\hat{\Omega}_c^2+\omega_+^2(\textbf q)\omega_-^2(\textbf q)(\hat{\mathbb1}+\alpha\hat{\mathcal{K}}^2)^{-1}\big)\hat{\mathcal{K}}_c\bigg].
\end{align}
\begin{figure*}[t!]
    \centering
    \includegraphics[width=1.0\textwidth,keepaspectratio]{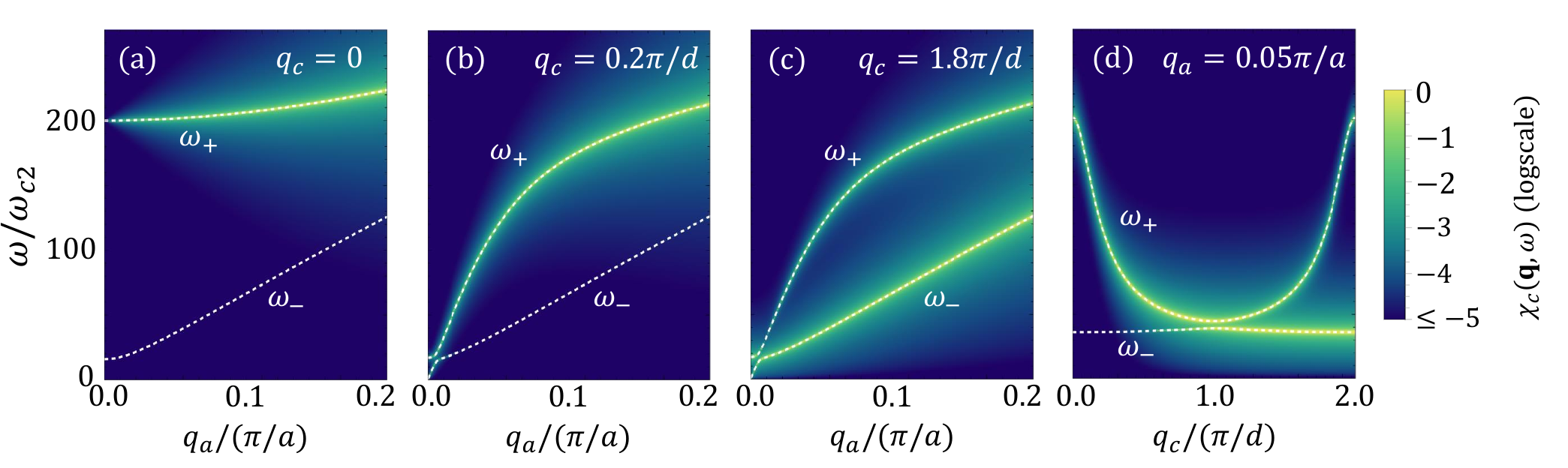}
    \caption{Intensity map of $\chi_c(\textbf q,\omega)$ of a bilayer superconductor according to Eq.\ \eqref{measquant} as a function of $q_a$ for fixed (a) $q_c=0$, (b) $q_c=0.2\pi/d$, (c) $q_c=1.8\pi/d$, or (d) as a function of $q_c$ for fixed $q_a=0.05\pi/a$. Every intensity map is normalized to its maximum value. White dashed lines represent the RPA dispersions $\omega_+(\textbf q)$ and $\omega_-(\textbf q)$. In the plot we used $\omega_{ab}/\omega_{c2}=200$, $\omega_{c1}/\omega_{c2}=12$, $d_1/d=0.3$, $a/d=0.35$, $\alpha/d^2=0.08$ and $\gamma/\omega_{c2}=0.05$.}
    \lb{fig3}
\end{figure*}\\
One can thus clearly see that in principle both plasmons appear as poles of the density response. After the analytic continuation $i\Omega_m\to\omega+i0^+$, the full charge response reads \cite{hepting_rixs_prb24,abbamonte_nature23}
\begin{align}\lb{measquant}
    &\chi_c(\textbf q,\omega)=-\text{Im}\sum_{\alpha,\beta}[\hat\chi_{\rho\rho}]_{\alpha\beta}=\nn\\
    &W_+(\textbf q)\delta\big(\omega-\omega_+(\textbf q)\big)+W_-(\textbf q)\delta\big(\omega-\omega_-(\textbf q)\big),
\end{align}
where $W_{\pm}(\textbf q)$ are the spectral weights of the two modes. This quantity is shown in Fig.\ \ref{fig3}, where we introduced a phenomenological damping parameter $\gamma$.
To have a general understanding of the spectral weights we can consider the limit of infinite compressibility, $\alpha\to0$, which does not change significantly the overall behavior of the response function and allows us to write $W_{\pm}(\textbf q)$ in a compact way. In this case one finds
\begin{align}\lb{W+}
    &W_+(\textbf q)=\frac{\varepsilon_\infty}{2}\frac{\omega_+(\textbf q)}{\omega_+^2(\textbf q)-\omega_-^2(\textbf q)}\bigg[\frac{\sin^2(\frac{q_a a}{2})}{a^2}\big(\omega_{ab}^2-\omega_-^2(\textbf q)\big)\nn\\
    &+\frac{\sin^2(\frac{q_cd_1}{2})}{dd_1}\big(\omega_{c1}^2-\omega_-^2(\textbf q)\big)+\frac{\sin^2(\frac{q_cd_2}{2})}{dd_2}\big(\omega_{c2}^2-\omega_-^2(\textbf q)\big)\bigg]
\end{align}
and
\begin{align}\lb{W-}
    &W_-(\textbf q)=\frac{\varepsilon_\infty}{2}\frac{\omega_-(\textbf q)}{\omega_+^2(\textbf q)-\omega_-^2(\textbf q)}\bigg[\frac{\sin^2(\frac{q_a a}{2})}{a^2}\big(\omega_+^2(\textbf q)-\omega_{ab}^2\big)\nn\\
    &+\frac{\sin^2(\frac{q_cd_1}{2})}{dd_1}\big(\omega_+^2(\textbf q)-\omega_{c1}^2\big)+\frac{\sin^2(\frac{q_cd_2}{2})}{dd_2}\big(\omega_+^2(\textbf q)-\omega_{c2}^2\big)\bigg].
\end{align}
One can check that $W_+(\textbf q)$ is nonvanishing for any finite $\textbf q$. On the contrary, one finds that $W_-(\textbf q)=0$ for $q_c=0$, due to two concomitant effects in Eq.\ \eqref{W-}: (i) in the first row, the frequency of the upper mode coincides with the in-plane plasma frequency, $\omega_+(q_a,0)=\omega_{ab}$, so this term vanishes; (ii) the terms in the second row scale as $\sin(q_c d_\lambda/2)$, which vanishes for $q_c=0$. One then recovers that the lower plasmon disappears from the density-density response at $q_c=0$ for any value of $q_a$, while its spectral weight grows for $q_c$ approaching the Brillouin zone boundary, see Fig.\ \ref{fig3}(d). Interestingly, we find that while the dispersions of the Josephson plasmons are $q_c$-periodic, the spectral weights of the density-density response associated with the modes are not, and in particular the lower plasmon has finite spectral weight for $q_c=2\pi/d$.\\
Our results are consistent with previous derivations based on the RPA calculation of the density response in the metal \cite{hepting_rixs_prb24,stoof_prb24,yamase_cm24}. As mentioned above, such agreement is expected as our approach is formally equivalent to the standard RPA one. However, here we can not only provide an analytical expression for the density response in the SC state Eq.\ \eqref{measquant}, but also a physical explanation of its behavior in terms of the polarization of the modes, as discussed in the next Section.
\subsection{Polarizations of the plasma modes}
The presence of a ghost plasmon in the density-density response computed in the previous Section is apparently rather unexpected. Indeed, as we mentioned in the introduction, plasmons appear in the density response because they identify the \textit{longitudinal} components of the electric field $\textbf E$ \cite{maier}. Their link to the charge density $\rho$ is encoded in the first Maxwell's equation, $\boldsymbol\nabla\cdot\textbf E=4\pi\rho$, which connects $\rho$ to the longitudinal component of the electric field $\text E_L\equiv \textbf q \cdot \textbf E$.  This is the reason why one expects, on physical grounds, to see signatures of a plasma mode in the density response. At the same time, the continuity equation $\partial_t\rho+\boldsymbol{\nabla}\cdot\textbf J=0$  links the density fluctuations to the longitudinal component of the current $\textbf J$. Since the latter is proportional to the longitudinal component of the electric field $\textbf E$, the two conditions are equivalent. In addition, for isotropic systems in the SC state the phase degrees of freedom in the Coulomb gauge $\textbf q\cdot \textbf A=0$ are decoupled from the transverse e.m.\ field, see discussion below Eq.\ \pref{coupactSL}, and the modes described by the phase fluctuations are automatically the longitudinal plasmons. As a consequence, the poles of the phase propagator coincide with the poles of the density response.\\
However, when the spatial isotropy is broken, as it occurs in layered systems with different in-plane and out-of-plane plasma frequencies, several additional effects arise, as recently discussed in the context of layered superconductors \cite{gabriele_prr22,sellati_prb23} and metals \cite{gabriele_prb24}. The first one is that the two e.m.\ modes, i.e.\ the longitudinal plasmon and the transverse polariton, are in general \textit{mixed} at finite momenta. This effect can be included in the present formalism by retaining the coupling between $\theta$ and $\textbf A$ in Eq.\ \pref{coupactSL}, that is equivalent to include retardation effects for the electric fields. Nonetheless, as discussed in Refs.\ \cite{gabriele_prr22,sellati_prb23, gabriele_prb24}, for what concerns the plasmon dispersion this effect becomes quantitatively relevant only below a momentum scale $\bar q\sim \sqrt{\omega_{ab}^2-\omega_c^2}/c$, which is typically of the order of $\mu\text{m}^{-1}$ and thus much smaller than those accessible by RIXS or EELS. In other words, for $|\textbf q| \gg \bar q$ the e.m.\ modes are purely transverse or longitudinal, and then the approximation used so far by including only the coupling of $\theta$ to the scalar potential $\phi$ to derive the longitudinal plasmons is quantitatively good. The crucial point, that is hidden in the derivation of the previous Section, is that in the bilayer system the lower Josephson plasmon is longitudinal with respect to a "reminiscent" large $q_c$, and as such it is always polarized in the out-of-plane direction, regardless of the direction of the momentum. Since the density fluctuations pick up only longitudinal currents, they do not see the lower plasmon until the momentum $q_c$ is big enough to give it a consistent longitudinal component. \\
A simple argument to explain this effect is provided in Fig.\ \ref{fig4}. As we have seen before, in a single-layer superconductor with Josephson coupling constant $J$ the dispersion of the unique plasma mode at finite momentum is proportional to a weighted average of the two scales $\omega_{ab}$ and $\omega_c$, as given by Eqs.\ \eqref{RPAdispSL2} and \eqref{RPAdispSL}. In particular, if one fixes the in-plane component of the momentum $q_a$ (considered small but finite in the figure, $q_a\sim 0$) and varies $q_c$, the dispersion of the plasmon crosses over from the scales of the large in-plane frequency $\omega_{ab}$, that are realized when $q_c=0$, to the scales of the small out-of-plane frequency $\omega_c$, that characterize the longitudinal mode when the momentum has $q_c=\pi/d$, with $d$ representing here the single-layer periodicity in the $c$-direction. This dispersion is shown in Fig.\ \ref{fig4}(a). In the bilayer crystal one introduces an additional breaking of the translational symmetry in the $c$-direction due to the fact that the Josephson couplings are different, $J_1\neq J_2$. As a consequence, the new Brillouin zone in the $c$-direction has half the original size, and the lower part of the single-layer plasmon dispersion is folded back to $q_c=0$, see Fig.\ \ref{fig4}(b). The lower plasmon of the bilayer system can then be understood as an image of the backfolded original plasmon, and as such it retains a memory of the large out-of-plane momentum it had at the Brillouin-zone boundary of the single-layer cell. This affects both its polarization, that is connected to currents that are perpendicular to the planes, and the staggered, counterflowing pattern of the currents themselves. When one accounts for the further splitting of the lattice spacings by setting $d_1\neq d_2$, the degeneracy at the boundary of the bilayer Brillouin zone is lifted, and the folded lower plasmon splits from the upper plasmon branch, see Fig.\ \ref{fig4}(c). The latter essentially behaves as the unique plasmon of the single-layer superconductor.
\begin{figure}[t!]
    \centering
    \includegraphics[width=0.49\textwidth,keepaspectratio]{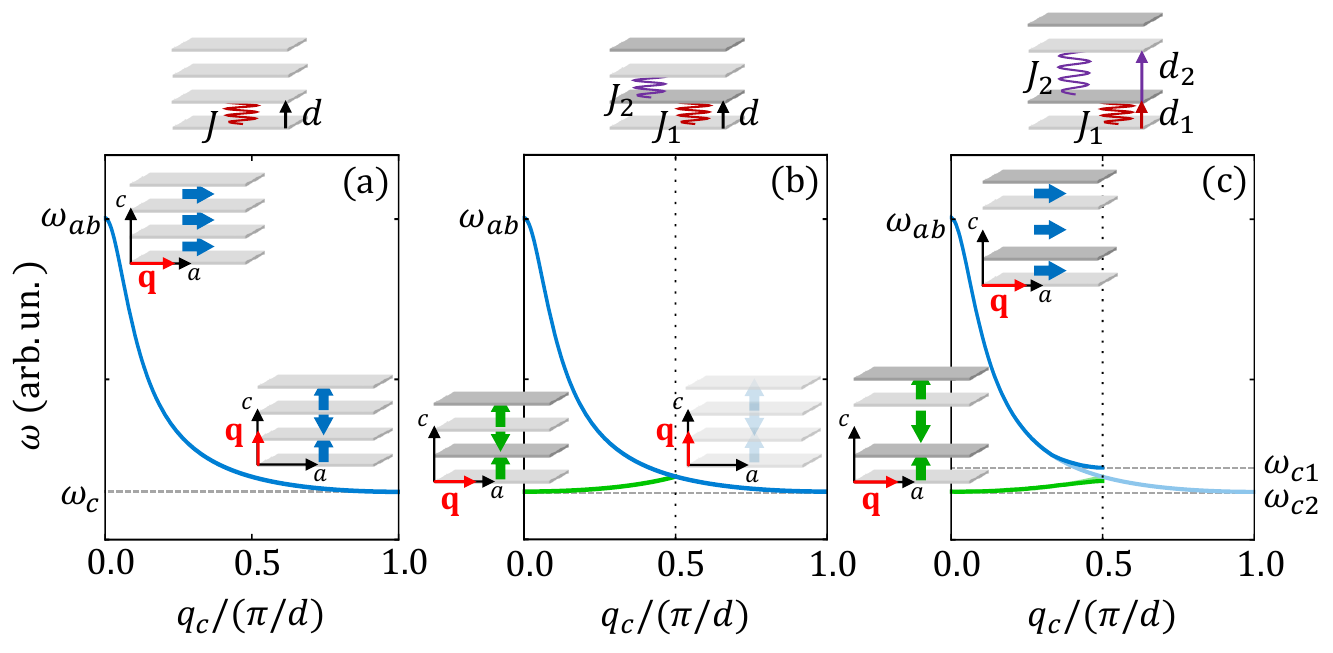}
    \caption{Sketch of the backfolding mechanism that leads to the two Josephson plasmons in the bilayer system. (a) Dispersion on the longitudinal plasmon in a single-layer system for $q_a\sim 0$ as a function of $q_c$, according to Eq.\ \pref{RPAdispSL2}. (b) When one makes the Josephson coupling among layers inequivalent the system has double periodicity in the $c$-direction, that implies a halving of the Brillouin zone in momentum space. The lower part of the plasmon dispersion is then backfolded to $q_c=0$, leading to a plasmon mode that in the new scheme appears with a {\em transverse staggered} polarization, as it was in the original single-layer case. (c) When one further accounts for the lattice anisotropy $d_1\neq d_2$ a gap opens at the boundary of the new Brillouin-zone, that corresponds to the splitting of the plasmons in the small $q_a$ limit for the bilayer case. }
    \lb{fig4}
\end{figure}\\
A convenient and efficient way to formally compute explicitly the polarizations of the Josephson modes is to fix the Weyl gauge $\boldsymbol{\phi}=0$ and introduce the gauge-invariant combination of the phase and e.m.\ vector potential, $\boldsymbol\psi_a(q)=(\psi_{a1}(q),\psi_{a2}(q))^T$ and $\boldsymbol\psi_c(q)=(\psi_{c1}(q),\psi_{c2}(q))^T$, defined as
\begin{align}\lb{gaugeinvBL}
        \boldsymbol\psi_a(q)= ik_a\boldsymbol\theta(q)+\frac{2e}{c}\textbf A_{a}(q),\nn\\
    \boldsymbol\psi_c(q)= i\hat{\mathcal{K}}_c\boldsymbol\theta(q)+\frac{2e}{c}\textbf A_{c}(q),
\end{align}
where $\textbf A_{a}(q)=(\text A_{a1}(q),\text A_{a2}(q))^T$ represent the in-plane and $\textbf A_{c}(q)=(\sqrt{2d_1/d}\text{A}_{c1}(q),\sqrt{2d_2/d}\text{A}_{c2}(q))^T$ represent the out-of-plane components of the vector potential. The technical details of this formalism are provided in Appendix C. Within this framework, the action that couples the SC phase to the e.m.\ field is expressed as $S[\boldsymbol\theta,\boldsymbol\psi_a,\boldsymbol\psi_c]$, and by integrating out $\boldsymbol \theta(q)$ one recovers a complete description of the collective modes in terms of the fields $\boldsymbol\psi_{a,c}$ \cite{sellati_prb23} which, according to their definitions Eq.\ \eqref{gaugeinvBL}, are formally proportional to physical currents. 
In particular, $\psi_{a\lambda}$ encases information on the in-plane oscillations in the $\lambda$th layer, whereas $\psi_{c1}$ and $\psi_{c2}$ describe respectively intrabilayer and interbilayer out-of-plane oscillations. The polarizations of the Josephson plasmons can thus be studied in terms of the components $\boldsymbol{\psi}_a^\pm(\textbf q)$ and $\boldsymbol{\psi}_c^\pm(\textbf q)$ of their associated eigenvectors. Their general behavior is sketched in Fig.\ \ref{fig5}(a,b).
To quantify the longitudinal projection of the current fluctuations associated with the plasma modes we define the normalized quantity
%
\begin{equation}\lb{projections}
\psi_L^{\pm}(\textbf q)=\frac{q_a|\boldsymbol\psi^\pm_{a}(\textbf q)|+q_c|\boldsymbol\psi^\pm_{c}(\textbf q)|}{|\textbf q|},
\end{equation}
\begin{figure}[t!]
    \centering
    \includegraphics[width=0.49\textwidth,keepaspectratio]{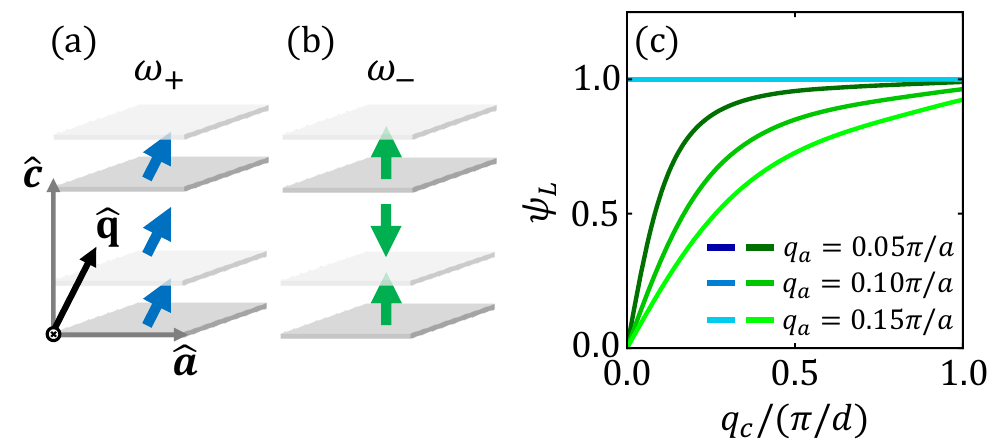}
    \caption{(a) Schematic representation of the polarization of the upper Josephson mode $\omega_+(\textbf q)$. The currents, represented by blue arrows, are along the direction of the momentum $\textbf q$ and oscillate in phase between neighbouring layers. (b) Schematic representation of the polarization of the lower Josephson mode $\omega_-(\textbf q)$. The currents, represented by green arrows, are along the $c$-axis for any direction of $\textbf q$ and their oscillations in the intrabilayer and interbilayer spacings are in opposite phase. In both panels, the length and width of an arrow do not carry any information on its magnitude. (c) Longitudinal projection of the polarizations associated with the upper $\psi^+_L(\textbf q)$ (blue) and lower $\psi^-_L(\textbf q)$ (green) Josephson plasmons according to Eq.\ \eqref{projections}, as function of $q_c$ for fixed $q_a$.}
    \lb{fig5}
\end{figure}
which is shown in Fig.\ \ref{fig5}(c). As one expects for a typical plasmon, the upper mode is longitudinal regardless of the direction of the momentum, so that $\psi^+_L(\textbf q)=1$ for every value of $\textbf q$. Moreover, the components of its eigenvector are such that $\psi_{a1}^+=\psi_{a2}^+$ and $\psi_{c1}^+/\psi_{c2}^+>0$, corresponding to in-phase oscillations along the planes and in the out-of-plane direction, see Fig.\ \ref{fig5}(a). On the other hand, the lower Josephson plasmon is a \textit{transverse} mode at small $q_c$: as we discussed above, the current fluctuations associated with this mode retain the original polarization along $c$ of the single-layer unit cell, such that only for finite $q_c$ the longitudinal projection $\psi_L^-(\textbf q)$ is sizable. Even so, the components of the eigenvector of the lower plasmon are such that $\psi_{c1}^-/\psi_{c2}^-<0$, so that the mode is associated with counterflowing currents in the intrabilayer and interbilayer spacings, see Fig.\ \ref{fig5}(b), although their absolute value is generically different and strongly dependent on the lattice parameters of the system. The counterflowing currents weaken the coupling of the density probe to the mode, which is then visible only for large $q_c$ where the longitudinal projection is $\psi_L^-(\textbf q)\sim 1$, as one can observe in Fig.\ \ref{fig3}(d). To further verify that the reason why the lower Josephson plasmon behaves as a ghost is because it is $c$-axis polarized and not a neutral mode, we show explicitly in Fig.\ \ref{fig6} the separate density fluctuations in each layer, defined as 
$\chi_{c}^{(1)}(\textbf q,\omega)=-\text{Im}\big[[\hat\chi_{\rho\rho}]_{11}+[\hat\chi_{\rho\rho}]_{12}\big]$ and $\chi_{c}^{(2)}(\textbf q,\omega)=-\text{Im}\big[[\hat\chi_{\rho\rho}]_{21}+[\hat\chi_{\rho\rho}]_{22}\big]$. As one can see in Fig.\ \ref{fig6}(a,b), the density response at the lower plasmon branch separately vanishes in each layer for small $q_c$, due to the fact that the current fluctuations are transverse and do not induce density fluctuations. On the other hand for large $q_c$, see Fig.\ \ref{fig6}(c,d), the currents associated with the lower plasmon gain a finite longitudinal component and signatures of its dispersion appear in both density fluctuations $\chi_\text{ch}^{(1)}$ and $\chi_\text{ch}^{(2)}$, with opposite signs but with different intensities indicating counterflowing but not canceling currents.
%
\begin{figure}[t!]
    \centering
    \includegraphics[width=0.49\textwidth,keepaspectratio]{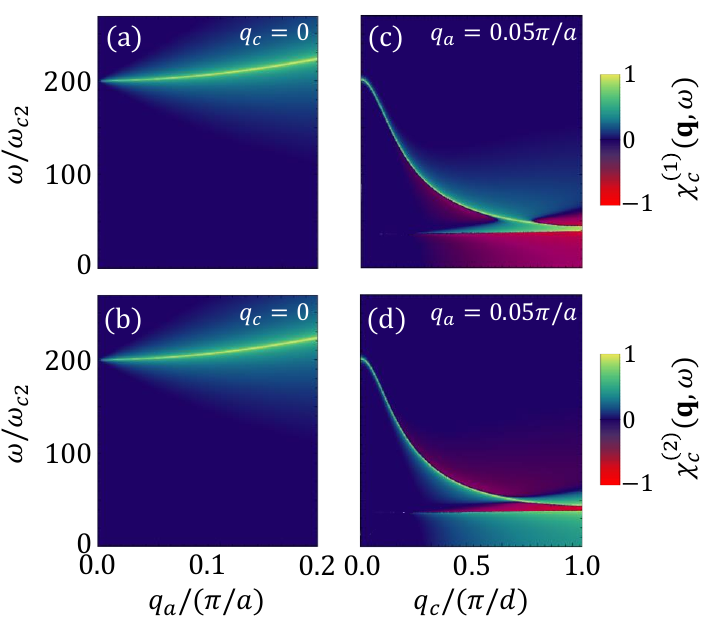}
    \caption{Intensity map of $\chi_c^{(\lambda)}(\textbf q,\omega)$ for (a,c) $\lambda=1$ and (b,d) $\lambda=2$. In the plots we used the same parameters as in Fig.\ \ref{fig3}. Color bars are in logarithmic scale, saturated with blue for values between $-10^{-5}$ and $10^{-5}$ and normalized to the maximum absolute value.}
    \lb{fig6}
\end{figure}
\subsection{Comparison with RIXS experiment on Ca-YBCO}
We now compare our results with experimental RIXS data from Ref.\ \cite{hepting_rixs_prb24} on Ca-YBCO, with fixed out-of-plane momentum $q_c=1.8\pi/d$. As we discussed above, the density-density response function is not a periodic function of $q_c$, so that the intensity of the two branches for $q_c=1.8\pi/d$ differs from that at $q_c=-0.2\pi/d$ and thus, by symmetry, from that at $q_c=0.2\pi/d$. In particular, the lower Josephson plasmon is visible in the former case, as one can see comparing Figs.\ \ref{fig3}(b) and \ref{fig3}(c). This effect is easily understood within the physical interpretation proposed above, as the longitudinal projection of the lower mode is much bigger for $q_c=1.8\pi/d$ than $q_c=0.2\pi/d$. We thus argue that the mode observed in Ref.\ \cite{hepting_rixs_prb24} is the lower plasmon, in agreement with Ref.\ \cite{yamase_cm24}, while the upper plasmon, falling well within the quasiparticle continuum, is overdamped \cite{abbamonte_natcomm23}. In Fig.\ \ref{fig7} we show the experimental points plotted over the theoretical density-density response according to Eq.\ \eqref{measquant}. The lower branch fits remarkably well the experiment, with reasonable fit parameters for a cuprate. We note that it would still be possible to fit the experimental points with the upper branch, as originally suggested in Ref.\ \cite{hepting_rixs_prb24} but with the correct out-of-plane momentum $q_c=1.8\pi/d$. However, this would require to use an in-plane plasma frequency in the range of $\omega_{ab}\sim 0.2-0.5\text{ meV}$, that seems smaller than what one would expect from in-plane electronic transport measurements with EELS \cite{fink_prb89,fink_prb91,fink_cm24} or RIXS \cite{lee_rixs_nature18,liu_rixs_npjqm20} in other families of cuprates. Even if this was the case, we find that signatures of the lower branch would be visible, emerging from below the RIXS elastic peak at least for the spectra at $q_a=0.08\pi/a$ and $q_a=0.10\pi/a$. As such, it seems to us more plausible to assign the measured dispersion to the lower plasmon branch.
\begin{figure}[t!]
    \centering
    \includegraphics[width=0.49\textwidth,keepaspectratio]{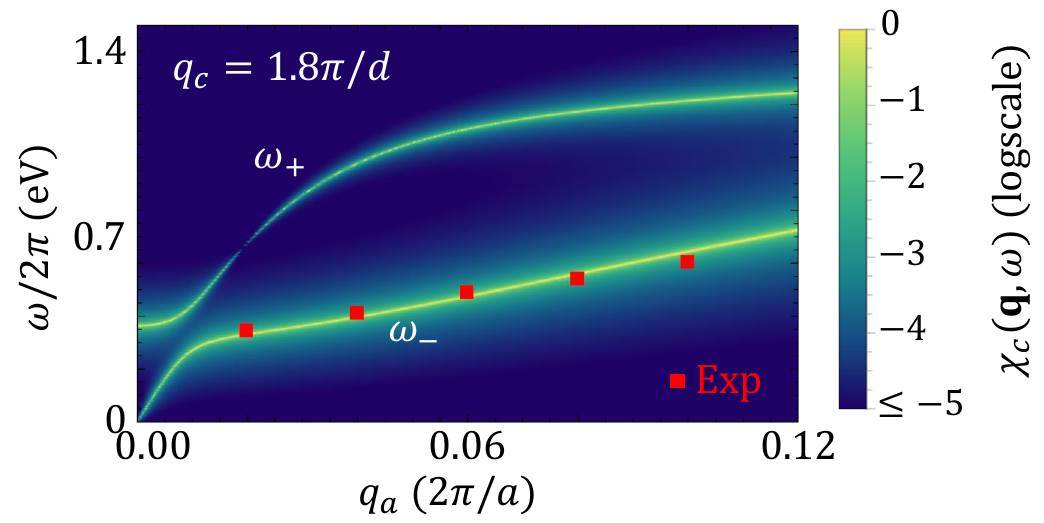}
    \caption{Comparison between $\chi_c(\textbf q,\omega)$ (color map) as in Eq.\ \eqref{measquant} and experimental points (red squares) extracted from Ref.\ \cite{hepting_rixs_prb24}. The response function is normalized to its maximum value. Lattice parameters are $a=3.83\text{ }\angstrom$, $d_1=3.37\text{ }\angstrom$, $d=11.74\text{ }\angstrom$ \cite{singh_ssc07}. Fit parameters are $\omega_{ab}/2\pi=1.24$ eV, $\omega_{c1}/2\pi=330$ meV, $\omega_{c2}/2\pi=4.96$ meV, $\alpha=2.07\text{ }\angstrom^2$ and $\gamma/2\pi=0.248$ meV.}
    \lb{fig7}
\end{figure}
\section{Conclusions}
In this manuscript we provided a detailed analysis of the density-density response in bilayer superconductors, systems with two SC sheets per unit cell and anisotropic Josephson-like couplings between neighboring planes belonging to the same unit cell, or two adjacent ones. We obtain an analytical expression for the density-density response function in the SC state, which shows a vanishing of the spectral weight of the lower Josephson plasmon as the out-of-plane momentum $q_c$ goes to zero. Such an unexpected behavior can be understood by studying the polarization of the corresponding current fluctuations, that we can compute by taking advantage of a compact formulation by means of the gauge-invariant SC currents. We then find that the lower plasma mode corresponds to staggered current fluctuations polarized perpendicularly to the planes. On physical ground, this effect can be understood as a consequence of the backfolding of the single-layer plasmon dispersion at the $q_c$ zone boundary to $q_c=0$: the lower plasmon then appears as a mode polarized longitudinally to a virtually large reminiscent momentum $q_c$, and it becomes a ghost to the physical density response for a small experimental $q_c$. Interestingly, in the region of large $q_c$ in which the mode becomes visible in the density response, the density fluctuations in the two layers of each unit cell have opposite signs and the dispersion of the mode becomes linear, resembling a sound rather than a flat, gapped plasmon. This is the same behavior observed for the sound-like demon mode emerging in multiband metallic $\text{Sr}_2\text{RuO}_4$ in the theoretical calculations of Ref.\ \cite{abbamonte_nature23}. Whether or not this is a general feature in multicomponent metals and superconductors remains an interesting open question, to be explored in future work.

\vspace{1cm} {\bf Acknowledgments}
We acknowledge financial support by EU under project MORE-TEM ERC-Syn grant agreement No.\ 951215 and by Sapienza University of Rome under Project Ateneo 2022 RP1221816662A977 and Project Ateneo 2023 RM123188E357C540.

\appendix

\section{Density-density response function in the effective-action formalism}
In this Appendix we provide the main steps to compute the density-density response function in single-layer superconductors. In the following we will adopt a path-integral formalism, that proves effective in describing plasma modes in the SC state by means of the phase fluctuations $\theta$. 
The Gaussian action that describes the bare phase fluctuations can be written as the anisotropic generalization of the phase-only action of isotropic superconductors \cite{randeria_prb00,benfatto_prb01,depalo_prb99,benfatto_prb04}, $S_G^\text{iso}[\theta]=1/8 \sum_q[\kappa_0\Omega_m^2+D_s|\textbf k|^2]|\theta(q)|^2$ with $D_s$ the isotropic superfluid stiffness, that describes a sound-like mode. In single-layer crystals this can be written as Eq.\ \eqref{sthetaSL} \cite{gabriele_prr22,benfatto_prb04}, which can be derived by assuming a Josephson-like interaction between subsequent planes along the $c$-axis \cite{sellati_prb23,sellati_nano24}. One can then introduce the e.m.\ field and the external probe through the minimal coupling substitution $\Omega_m\theta(q)\to\Omega_m\theta(q)-2e\phi(q)-2e\delta\phi(q)$. Adding the free e.m.\ action Eq.\ \eqref{freeemSL} the total action of the system reads
\begin{align}
    S_\text{RPA}[&\theta,\phi,\delta\phi]=S_\text{RPA}[\theta,\phi]-\frac{\varepsilon_\infty d}{8\pi}\sum_q k_\text{TF}^2|\delta\phi(q)|^2\nn\\
    -&\frac{\varepsilon_\infty d}{8\pi}\sum_q \big[k_\text{TF}^2\big(\phi(q)\delta\phi(-q)+\phi(-q)\delta\phi(q)\big)\big]\nn\\
    +&\frac{ed}{4}\sum_q\big[\kappa_0\Omega_m\big(\theta(q)\delta\phi(-q)-\theta(-q)\delta\phi(q)\big)\big],
\end{align}
with $S_\text{RPA}[\theta,\phi]$ defined in Eq.\ \eqref{RPAactSL} of the main text. As we discussed in Section II.A, the integration of the scalar potential $\phi$ dresses the phase mode, promoting it to a plasmon. One consequently gets 
\begin{align}
    &S_\text{RPA}[\theta,\delta\phi]=S_\text{RPA}[\theta]-\frac{\varepsilon_\infty d}{8\pi}\sum_q \frac{|\textbf k|^2}{1+\alpha|\textbf k|^2}|\delta\phi(q)|^2\nn\\
    &+\frac{\varepsilon_\infty d}{16\pi e}\sum_q \frac{\Omega_m|\textbf k|^2}{1+\alpha|\textbf k|^2}(\theta(q)\delta\phi(-q)-\theta(-q)\delta\phi(q)),
\end{align}
where $S_\text{RPA}[\theta]$ defined in Eq.\ \eqref{RPAdressSL} of the main text. To compute the density-density response function one should then integrate out $\theta$, resulting in the effective action
\begin{align}\lb{densdensactSL1}
    S_\text{RPA}[\delta\phi]&=-\frac{\varepsilon_\infty d}{8\pi}\sum_q \frac{|\textbf k|^2}{1+\alpha|\textbf k|^2}\nn\\
    &\times\bigg[1-\frac{\Omega_m^2|\textbf k|^2}{1+\alpha|\textbf k|^2}\langle\theta(q)\theta(-q)\rangle\bigg]|\delta\phi(q)|^2,
\end{align}
where
\begin{align}
 \langle\theta(q)\theta(-q)\rangle=\bigg[\frac{|\textbf k|^2}{1+\alpha|\textbf k|^2}\big(\Omega_m^2+\omega_\text{RPA}^2(\textbf q)\big)\bigg]^{-1}   
\end{align}
is the plasmon propagator, defined as the inverse of the coefficient of $|\theta(q)|^2$ in $S_\text{RPA}[\theta]$ in Eq.\ \eqref{RPAdressSL}. One can see from Eq.\ \eqref{densdensactSL1} that the matter degrees of freedom $\theta$, i.e.\ the plasmon, is dressing the probe field to give the density response. Explicitly, the effective action reads  
\begin{align}\lb{densdensactSL}
    S_\text{RPA}[\delta\phi]=-\frac{\varepsilon_\infty d}{8\pi}\sum_q\frac{|\textbf k|^2\omega_L^2(\textbf q)}{\Omega_m^2+\omega_\text{RPA}^2(\textbf q)}|\delta\phi(q)|^2,
\end{align}
from which one can find the density-density response function Eq.\ \eqref{densdensfuncSL} in the main text through its definition Eq.\ \eqref{densdensdef}.
\section{Phase-only description of Josephson plasmons in bilayer superconductors}
In this Appendix we provide the main steps to compute the RPA phase-only action and the density-density response function in the bilayer system. In this case the Josephson-like coupling between layers in the intrabilayer spacing differs from that in the interbilayer spacing, and one should thus define two different out-of-plane superfluid stiffnesses $D_{c1}$ and $D_{c2}$. This immediately implies the existence of two Josephson plasma resonances, at frequencies $\omega_{c1}^2=4\pi e^2D_{c1}/\varepsilon_\infty$ and $\omega_{c2}^2=4\pi e^2D_{c2}/\varepsilon_\infty$. Moreover, as discussed in the main text, the phases of the two SC planes are generically different and thus we define the basis $\boldsymbol{\theta}(q)=\big(\theta_1(q),\theta_2(q)\big)^T$, where $\theta_\lambda$ refers to the fluctuations in the $\lambda$th layer. With all these prescriptions, Eq.\ \eqref{sthetaSL} that describes the bare phase fluctuations can be written in the bilayer system as \cite{sellati_prb23,fiore_prb24}
\begin{align}\lb{sthetaBLapp}
    S_G[\boldsymbol\theta]&=\frac{\varepsilon_\infty}{32\pi e^2}\frac{d}{2}\sum_q\boldsymbol{\theta}^T(-q)\nn\\
    &\times\bigg[\frac{\Omega_m^2}{\alpha}\hat{\mathbb1}+\omega_{ab}^2k_{ab}^2\hat{\mathbb1}+\hat{\mathcal{K}}_c^\dagger\hat{\Omega}_c^2\hat{\mathcal{K}}_c\bigg]\boldsymbol{\theta}(q),
\end{align}
where we denote $2\times 2$ matrices with the hat symbol. Following the procedure detailed above for the single-layer superconductor, we now introduce the e.m.\ field in the system. We note that in principle the generalization of the minimal-coupling substitution and of Maxwell's equations to the bilayer system is not straightforward, and requires careful definition of the fields on the lattice and of the Fourier transform on the discrete $c$-direction \cite{homann_prr20,homann_prb21,sellati_prb23}.
As we are interested in describing the density-density response in the nonrelativistic regime, we again study the plasma modes neglecting retardation effects and thus only introducing the scalar potential $\boldsymbol{\phi}(q)=\big(\phi_1(q),\phi_2(q)\big)^T$. The action for the free e.m.\ scalar potential in the bilayer systems can be written as the generalization of Eq.\ \eqref{freeemSL}, and it reads \cite{sellati_prb23}
\begin{align}\lb{freeemBLapp}
    S_\text{e.m.}[\boldsymbol\phi]=-\frac{\varepsilon_\infty}{8\pi}\frac{d}{2}\sum_q\boldsymbol\phi^T(-q)\big[k_a^2\hat{\mathbb1}+\hat{\mathcal{K}}_c^\dagger\hat{\mathcal{K}}_c\big]\boldsymbol\phi(q).
\end{align}
We then introduce the coupling between phase and scalar potential by performing the minimal-coupling substitution in Eq.\ \eqref{sthetaBLapp},
\begin{align}\lb{mincoupBL1app}
    \Omega_m\boldsymbol{\theta}(q)\to\Omega_m\boldsymbol\theta(q)-2e\boldsymbol\phi(q).
\end{align}
We can then write the total action of the system as
\begin{align}\lb{RPAactBL}
    &S_\text{RPA}[\boldsymbol\theta,\boldsymbol\phi]=S_G[\boldsymbol\theta]\nn\\
    &-\frac{\varepsilon_\infty}{8\pi}\frac{d}{2}\sum_q\boldsymbol\phi^T(-q)\bigg[\frac{1}{\alpha}\hat{\mathbb1}+ k_a^2\hat{\mathbb1}+\hat{\mathcal{K}}_c^\dagger\hat{\mathcal{K}}_c\bigg]\boldsymbol\phi(q)\nn\\
    &+\frac{\varepsilon_\infty}{16\pi e}\frac{d}{2}\sum_q\bigg[\frac{\Omega_m}{\alpha}\big(\boldsymbol\phi^T(-q)\boldsymbol{\theta}(q)-\boldsymbol{\theta}^T(-q)\boldsymbol\phi(q)\big)\bigg],
\end{align}
which is the equivalent of Eq.\ \eqref{RPAactSL} to the bilayer case. To dress the phase fluctuations with the Coulomb interaction we then integrate out the scalar potential, and find the action that describes the two Josephson plasmons, Eq.\ \eqref{RPAdressBL} in the main text.\\
To compute the density-density response function we introduce the external density probe as an auxiliary field $\delta\boldsymbol{\phi}(q)=\big(\delta\phi_1(q),\delta\phi_2(q)\big)^T$ through the minimal-coupling substitution Eq.\ \eqref{mincoupBL1app}, that now reads
\begin{align}\lb{mincoupBL2app}    
\Omega_m\boldsymbol\theta(q)\to\Omega_m\boldsymbol\theta(q)-2e\boldsymbol\phi(q)-2e\delta\boldsymbol{\phi}(q).
\end{align}
Thus, the total action of the system Eq.\ \eqref{RPAactBL} has additional terms that couple $\delta\boldsymbol{\phi}$ to $\boldsymbol{\theta}$ and $\boldsymbol{\phi}$, and reads
\begin{align}
    &S_\text{RPA}[\boldsymbol{\theta},\boldsymbol{\phi},\delta\boldsymbol\phi]=S_\text{RPA}[\boldsymbol{\theta},\boldsymbol\phi]\nn\\
    &-\frac{\varepsilon_\infty}{8\pi}\frac{d}{2}\sum_q\delta\boldsymbol\phi^T(-q)\bigg(\frac{1}{\alpha}\hat{\mathbb1}\bigg)\delta\boldsymbol\phi(q)\nn\\
    &-\frac{\varepsilon_\infty}{8\pi}\frac{d}{2}\sum_q\bigg[\frac{1}{\alpha}\big(\delta\boldsymbol\phi^T(-q)\boldsymbol{\phi}(q)+\boldsymbol{\phi}^T(-q)\delta\boldsymbol\phi(q)\big)\bigg]\nn\\
    &+\frac{\varepsilon_\infty}{16\pi e}\frac{d}{2}\sum_q\bigg[\frac{\Omega_m}{\alpha}\big(\delta\boldsymbol\phi^T(-q)\boldsymbol{\theta}(q)-\boldsymbol{\theta}^T(-q)\delta\boldsymbol\phi(q)\big)\bigg].
\end{align}
Following the derivation of the single-layer case, we integrate out $\boldsymbol{\phi}$ to promote the phase modes to plasmons and then we integrate out $\boldsymbol\theta$ to find an effective action of the probe field $\delta\boldsymbol{\phi}$, that reads 
\begin{align}\lb{densdensactBLapp}
&S_\text{RPA}[\delta\boldsymbol\phi]=-\frac{\varepsilon_\infty}{8\pi}\frac{d}{2}\sum_q\delta\boldsymbol\phi^T(-q)\nn\\
&\times\bigg[\hat{\mathcal{K}}^2\big[\Omega_m^2\hat{\mathbb1}+\hat{\Omega}_L^2(\textbf q)(\hat{\mathbb1}+\alpha\hat{\mathcal{K}}^2)\big]^{-1}\hat{\Omega}_L^2(\textbf q)\bigg]\delta\boldsymbol\phi(q).
\end{align}
This action has a $2\times 2$ structure as it is written in the basis $\delta\boldsymbol{\phi}$. One can generalize the definition of the density-density response function Eq.\ \eqref{densdensdef} to the bilayer case as
\begin{align}\lb{densdensdefBLapp}
    \hat\chi_{\rho\rho}(q)=\frac{1}{d}\frac{\partial^2 S[\delta\boldsymbol\phi]}{\partial \delta\boldsymbol\phi(q)\partial \delta\boldsymbol\phi(-q)}.
\end{align}
By using Eq.\ \eqref{densdensactBLapp} in Eq.\ \eqref{densdensdefBLapp} one finally finds 
\begin{align}\lb{densdensfuncBL1app}
    \hat\chi_{\rho\rho}(q)=-\frac{\varepsilon_\infty}{8\pi}\hat{\mathcal{K}}^2\big[\Omega_m^2\hat{\mathbb1}+\hat{\Omega}_L^2(\textbf q)(\hat{\mathbb1}+\alpha\hat{\mathcal{K}}^2)\big]^{-1}\hat{\Omega}_L^2(\textbf q),      
\end{align}
which is Eq.\ \eqref{densdensfuncBL1} in the main text. One should take care that the system responds to the total density, i.e., the measured physical quantity is proportional to the sum over all the matrix elements of Eq.\ \eqref{densdensfuncBL1app}, as we wrote in Eq.\ \eqref{measquant}. 
\section{Gauge-invariant field description of plasma modes in bilayer superconductors}
In this appendix we apply the gauge-invariant field formalism to bilayer superconductors. We thus start back from Eq.\ \eqref{sthetaBLapp} and introduce the e.m.\ field, now fixing the Weyl gauge $\boldsymbol{\phi}=0$ and without neglecting the vector potential. In this case all the e.m.\ interactions are encoded in the latter, with in-plane components $\textbf A_{a}(q)=(\text A_{a1}(q),\text A_{a2}(q))^T$ and out-of-plane components $\textbf A_{c}(q)=(\sqrt{2d_1/d}\text{A}_{c1}(q),\sqrt{2d_2/d}\text{A}_{c2}(q))^T$. The latter are rescaled to obtain more compact expressions. The $b$-components of the e.m.\ vector potential gives two decoupled transverse polariton modes in the end \cite{sellati_prb23}, and we do not consider them here. With this gauge fixing, the free e.m.\ action Eq.\ \eqref{freeemSL} now reads \cite{sellati_prb23,homann_prb21,homann_prr20}
\begin{align}\lb{freeemBL2app}
    S_\text{e.m.}&[\textbf A]=\frac{\varepsilon_\infty}{8\pi c^2}\frac{d}{2}\sum_q\bigg[\textbf A_{c}^T(-q)\bigg(\Omega_m^2\hat{\mathbb1}+\frac{c^2}{\varepsilon_\infty}k_a^2\hat{\mathbb1}\bigg)\textbf A_{c}(q)\nn\\
    &+\textbf A_{a}^T(-q)\bigg(\Omega_m^2\hat{\mathbb1}+\frac{c^2}{\varepsilon_\infty}\hat{\mathcal K}_c^\dagger\hat{\mathcal K}_c\bigg)\textbf A_{a}(q)\nn\\
    &-\frac{c^2}{\varepsilon_\infty}k_a\big(\textbf A_{a}^T(-q)\hat{\mathcal K}_c^\dagger\textbf A_c(q)+\textbf A_{c}^T(-q)\hat{\mathcal K}_c\textbf A_{a}(q)\big)\bigg],
\end{align}
while the coupling of the phase modes to the vector potential is introduced in Eq.\ \eqref{sthetaBLapp} via the minimal-coupling substitution
\begin{align}\lb{mincoupBL3app}
    ik_a\boldsymbol\theta(q)\to ik_a\boldsymbol\theta(q)+\frac{2e}{c}\textbf A_{a}(q),\nn\\
    i\hat{\mathcal{K}}_c\boldsymbol\theta(q)\to i\hat{\mathcal{K}}_c\boldsymbol\theta(q)+\frac{2e}{c}\textbf A_{c}(q).
\end{align}
In this gauge the total action of the system $S[\boldsymbol\theta,\textbf A_{a},\textbf A_c]$ is a function of the vector quantities $\boldsymbol\theta(q)$, representing the matter degrees of freedom, and $\textbf A_{a,c}(q)$, representing the e.m.\ field. A very convenient way to study the collective plasma modes and their polarizations \cite{gabriele_prr22,sellati_prb23,gabriele_prb24} is to rescale the latter by introducing the 
gauge-invariant fields $\boldsymbol\psi_a(q)=(\psi_{a1}(q),\psi_{a2}(q))^T$ and $\boldsymbol\psi_c(q)=(\psi_{c1}(q),\psi_{c2}(q))^T$, defined as
\begin{align}\lb{gaugeinvBLapp}
        \boldsymbol\psi_a(q)= ik_a\boldsymbol\theta(q)+\frac{2e}{c}\textbf A_{a}(q),\nn\\
    \boldsymbol\psi_c(q)= i\hat{\mathcal{K}}_c\boldsymbol\theta(q)+\frac{2e}{c}\textbf A_{c}(q).
\end{align}
so that the action is expressed as $S[\boldsymbol\theta,\boldsymbol\psi_a,\boldsymbol\psi_c]$. By integrating out the matter degrees of freedom $\boldsymbol\theta$ one is left with
\begin{align}\lb{actgaugeBLapp}
    S[\boldsymbol\psi&]=\frac{\varepsilon_\infty}{32\pi e^2}\frac{d}{2}\sum_q\bigg[\nn\\
    +&\boldsymbol\psi_{a}^T(-q)\bigg(\Omega_m^2\hat{\mathbb1}+\omega_{ab}^2\hat{\mathbb1}+\frac{c^2}{\varepsilon_\infty}\hat{\mathcal K}_c^\dagger\hat{\mathcal K}_c\bigg)\boldsymbol\psi_a(q)\nn\\
    +&\boldsymbol\psi_{c}^T(-q)\bigg(\Omega_m^2\hat{\mathbb1}+\hat\Omega_c^2+\frac{c^2}{\varepsilon_\infty}k_a^2\hat{\mathbb1}\bigg)\boldsymbol\psi_c(q)\nn\\
    -&\frac{c^2}{\varepsilon_\infty}k_a\big(\boldsymbol\psi_a^T(-q)\hat{\mathcal{K}}_c^\dagger\boldsymbol{\psi}_c(q)+\boldsymbol\psi_c^T(-q)\hat{\mathcal{K}}_c\boldsymbol{\psi}_a(q)\big)\bigg].
\end{align}
In writing this expression we considered the limit for $\alpha\to0$ which is equivalent to taking infinite compressibility in the system. While in principle this is an approximation, it still gives a correct physical understanding of the behavior of the plasma oscillations, and it simplifies the problem as the polarizations can now be found as eigenvectors of the $4\times 4$ secular problem described by Eq.\ \eqref{actgaugeBLapp}. This action describes four mixed modes, two of which are the Josephson modes. Notice that the four modes cannot in principle be studied separately due to the coupling terms that mix $\boldsymbol{\psi}_a\boldsymbol{\psi}_c$. A thorough description of all the plasma modes of bilayer superconductors is given in Ref.\ \cite{sellati_prb23}, while in the main text we only focused on the polarizations of the two Josephson modes in the nonrelativistic regime, studying their eigenvectors $(\psi_{a1}^\pm,\psi_{a2}^\pm,\psi_{c1}^\pm,\psi_{c2}^\pm)$ numerically.

\bibliography{bibl.bib} 

\end{document}